\documentclass[12pt,a4paper]{article}
\usepackage{graphicx}
\usepackage{amsfonts}
\usepackage{cite}

\def\bea{\begin{eqnarray}}
\def\eea{\end{eqnarray}}
\def\bc{\begin{center}}
\def\ec{\end{center}}

\def\simlt{\stackrel{<}{{}_\sim}}
\def\simgt{\stackrel{>}{{}_\sim}}

\begin{document}
\pagestyle{empty}
\begin{flushright}
IFT-3/2012\\
{\tt hep-ph/yymmnnn}\\
{\bf \today}
\end{flushright}
\vspace*{5mm}
\begin{center}

{\large\bf Inert Dark Matter and
Strong Electroweak Phase Transition}\\
\vspace*{1cm}

{\bf Grzegorz Gil, Piotr Chankowski and Maria Krawczyk}

\vspace{0.5cm}

Faculty of Physics,  University of Warsaw, Ho\.za 69, 00-681,
Warsaw, Poland

\vspace*{1.7cm}
{\bf Abstract}
\end{center}
\vspace*{5mm}
\noindent
{The main virtue of the Inert  Doublet Model (IDM) is that one of its
spinless neutral bosons can play the role of Dark Matter.
Assuming that the additional sources of CP violation are present in
the form of higher dimensional operator(s) we reexamine
the possibility that the model parameters
for which the right number density of relic particles is predicted
are compatible with the first order phase transition that could lead
to electroweak baryogenesis.

We find, taking into account recent indications from the LHC and the
constraints from the electroweak precision data, that for a light DM 
(40-60 GeV) particle $H^0$ 
and heavy, almost degenerate additional scalars $H^\pm$ and $A^0$
this is indeed possible but the two parameters most important
for the strength of the phase transition: the common mass of $H^\pm$ and
$A^0$ and the trilinear coupling of the Higgs-like particle $h^0$ to DM
are strongly constrained.  $H^\pm$ and $A^0$ must weight less than
$\sim440$ GeV if the inert minimum is to be the lowest one
and the value of the $h^0H^0H^0$ coupling is limited by the XENON 100 data.
We  stress the important role of the zero temperature part of the
potential for the strength of the  phase transition.
}
\vspace*{1.0cm}
\date{\today}


\vspace*{0.2cm}

\vfill\eject
\newpage

\setcounter{page}{1}
\pagestyle{plain}

\section{Introduction}

If the  scalar sector of the theory of electroweak interactions consists
of only one  $SU(2)$ doublet, the electroweak $SU(2)_L\times U(1)_Y$ gauge
symmetry breaking down to $U(1)_{\rm EM}$ is the only possible pattern. The
first-order nature of the electroweak phase transition that had to occur
during the evolution of the Universe, as it cooled down below some critical
temperature $T_{\rm EW}$, becomes less and less pronounced with the increase
of the Higgs particle mass $M_h$ and turns into a continuous (second order)
transition above the mass $M_h\approx80$ GeV \cite{JANS}. With
the lower bound set on the mass of this particle by
searches at LEP and LHC \cite{LEP, LHC} the possibility
that the baryon asymmetry of the Universe was created during the
electroweak phase transition is definitely excluded within this
scenario, irrespectively of the question of amount of CP violation
predicted by the Standard Model to which in principle could contribute
also higher order nonrenormalizable operators \cite{TROD}.

The possibility of electroweak baryogenesis is still open in various
multiscalar extensions of the Standard Model which can exhibit more
complicated symmetry breaking patterns. In such theories spontaneous
breaking of CP is also possible in the scalar sector which can thus
contribute significantly to the CP violation necessary in cosmology.
An example worked out in detail is the electroweak baryogenesis in the
Minimal Supersymmetric Standard Model (MSSM) whose Higgs sector consists
of two scalar doublets. It has been shown \cite{WAGNERY} that in this model
the electroweak phase transition can remain sufficiently first order
even for the Higgs boson mass $\sim$ 100 GeV provided the right-chiral
top squark is light enough.
Tight conditions under which the phase transition predicted
by the MSSM is sufficiently first order are due to the  very constrained
form of the supersymmetric potential of the Higgs doublets and can be
relaxed in the nonsupersymmetric multiscalar models whose potentials
depend on many a priori free parameters.

Studies of the nature of the phase transition in the nonsupersymmetric
two Higgs doublet (2HDM) extension of the Standard Model were initiated
in the papers by Bochkaryev, Kuzmin and Shaposhnikov \cite{BOKUSH} and
of Turok and Zadrozny \cite{TURZAD}. Their authors, using the high
temperature approximation, have found that in this class of models
the rough criterion for the first-order phase transition
$v(T_{\rm EW})/T_{\rm EW}\simgt1$  can be
satisfied for the mass of the scalar playing the role
of the physical Higgs particle of order 100 GeV \cite{SCHAP}. The strength
of the electroweak phase transition predicted by the 2HDM was subsequently
analyzed in several papers \cite{LACAR,LASPZA,CLILEM,KAYASE,FROHUSE} which
using the one-loop temperature dependent effective potential (with or without
resummations of the so-called ring diagrams) have confirmed the original
observations. Recently, the interest in the 2HDM \cite{MA,BAHARY,BFLRSS},
and in the phase transition predicted by this model in particular
\cite{GIIVKA,GKKS,dsok,GIL,KOSKA,CLIKAITR}, has
again been revived.

There are many versions of the two scalar doublet extensions of the
Standard Model. A distinguished one which has recently attracted much
attention is the so-called Inert Doublet Model (IDM) possessing
an additional unbroken discrete $\mathbb{Z}_2$ symmetry \cite{MA,BAHARY}.
The main virtue of the IDM is that one of spin zero particles originating
from its scalar sector is stable and can play the role of the Dark Matter
\cite{laura,su}. Phase transitions in this model were  studied in
\cite{GKKS,dsok,GIL,CHNESEZH,BOCL}.

Of course, the unbroken $\mathbb{Z}_2$ symmetry of the IDM (crucial for
the stability of the dark matter particle) precludes any possibility of
additional CP violation originating from the scalar sector. Hence, in its
most attractive version, with a
stable dark matter candidate, the IDM predicts no more CP violation than
does the Standard Model. This means that the electroweak phase transition,
even if sufficiently first order, could not yield the required baryon number
asymmetry of the Universe. However if the IDM is viewed only as an effective
low energy theory (effectively parametrizing the electroweak symmetry
breaking) one can easily imagine that its full $\mathbb{Z}_2$-symmetric
Lagrangian involves also
CP violating nonrenormalizable terms \cite{DIHUSISU} like
\begin{eqnarray}
\Delta{\cal L}={1\over\Lambda^2}~\!(c_1\Phi^\dagger_1 \Phi_1+c_2\Phi^\dagger_2
\Phi_2)~\!{\rm tr}(W_{\mu\nu}\tilde W^{\mu\nu})\label{eqn:CPVOperator},
\end{eqnarray}
where $\Phi_i$ ($i=1,2$) are the two scalar doublets, $W_{\mu\nu}$ and
$\tilde W^{\mu\nu}$
the $SU(2)_L$ field strength and  its dual, $\Lambda$ the UV
cutoff and $c_i$ some coefficients.
Hence, the question of amount of CP violation in
the (renormalizable part of the) IDM might not be vital and the
question of electroweak baryogenesis hinges essentially only on the nature
of the electroweak phase transition.

The question whether the IDM parameters for which the right amount of dark
matter could be generated can also be compatible with the electroweak phase
transition of sufficiently first order to allow for electroweak baryogenesis
has been answered in the affirmative first in \cite{CHNESEZH} and quite
recently in \cite{BOCL}. In this letter we add a  few new points to these
analyses. Firstly,
using the full one-loop effective potential (with imposed physical
renormalization conditions, different than the ones used in \cite{BOCL}) we
show that the interesting configuration of the model parameters, with the DM
particle mass $\sim40\div80$ GeV and the mass $M_{h^0}$ of the physical
SM-like Higgs boson in the range 120$\div$130~GeV favoured by the LHC data
\cite{LHC}, is severely constrained by the requirement
that the IDM minimum is the lowest minimum. This puts upper limit
$\sim440$ GeV  on the masses of the additional heavy scalars
forcing them to be within the LHC reach. Secondly, by using the
full one-loop temperature dependent effective potential
(in \cite{CHNESEZH} only its high temperature expansion, without the
zero-temperature part was used) supplemented
with the resummation of the so-called ring diagrams we investigate
the strength of the electroweak phase transition. We find that
taking into account the zero-temperature part of the effective potential
has important effect on the phase transition making it significantly
stronger. However, similarly as the authors of  \cite{BOCL},
we find that the parameter space is rather
limited, especially if one includes the Xenon100 data \cite{XENON}.

\section{Parameters of the IDM}

We consider  a special case of the type I 2HDM in which only
one scalar doublet ($\Phi_1$) has Yukawa couplings to fermions. Owing to
this and to the following form of its Higgs potential
\begin{eqnarray}
V(\Phi_1,\Phi_2)=m^2_{11}\Phi_1^\dagger\Phi_1+m^2_{22}\Phi_2^\dagger\Phi_2
+{\lambda_1\over2}(\Phi_1^\dagger\Phi_1)^2
+{\lambda_2\over2}(\Phi_2^\dagger\Phi_2)^2\phantom{aaaaaaaaaaa}\nonumber\\
+\lambda_3(\Phi_1^\dagger\Phi_1)(\Phi_2^\dagger\Phi_2)
+\lambda_4(\Phi_1^\dagger\Phi_2)(\Phi_2^\dagger\Phi_1)
+{\lambda_5\over2}\left[(\Phi_1^\dagger\Phi_2)^2+(\Phi_2^\dagger\Phi_1)^2
\right]\label{eqn:2HDMpot}
\end{eqnarray}
the complete Lagrangian of the model possesses, in addition to
the gauge $SU(2)_L\times U(1)_Y$ symmetry, also the discrete
$\mathbb{Z}_2\times\mathbb{Z}_2^\prime$ symmetry:
\begin{eqnarray}
\mathbb{Z}_2:\phantom{a}\Phi_1\rightarrow\Phi_1~\!,\phantom{aaaaa}
\Phi_2\rightarrow-\Phi_2~\!,\phantom{aaaaaaaaaaaa}~
\nonumber\\
\mathbb{Z}_2^\prime:\phantom{a}\Phi_1\rightarrow-\Phi_1~\!,\phantom{aaa}
\phantom{a}\Phi_2\rightarrow\Phi_2~\!,\phantom{aaaa}f_R\rightarrow-f_R~\!,
\label{eqn:Z2symmetries}
\end{eqnarray}
($f_R$ denotes all right-chiral fermions; other SM fields are
$\mathbb{Z}_2\times\mathbb{Z}_2^\prime$-invariant).

All parameters of $V(\Phi_1,\Phi_2)$ are real. Depending on their values
different patterns of the
$SU(2)_L\times U(1)_Y\times\mathbb{Z}_2\times\mathbb{Z}_2^\prime$ symmetry
breaking are possible  - in some of them even the electromagnetic
$U(1)_{\rm EM}$ symmetry can be broken \cite{BAFESA,GIIVKA,GKKS}.
The tree-level potential for the doublets VEVs $v_1$ and $v_2$
defined by\footnote{In this paper we ignore possible
existence of $U(1)_{\rm EM}$ breaking  vacua.}
\begin{eqnarray}
\Phi_1={1\over\sqrt2}\left(\matrix{\sqrt2~\!G^+\cr v_1+h^0+iG^0}\right),
\phantom{aaaaa}
\Phi_2={1\over\sqrt2}\left(\matrix{\sqrt2~\!H^+\cr v_2+H^0+iA^0}\right)
\end{eqnarray}
reads
\begin{eqnarray}
V_{\rm tree}={1\over2}~\!m^2_{11}v^2_1+{1\over2}~\!m^2_{22}v^2_2
+{\lambda_1\over8}~\!v_1^4+{\lambda_2\over8}~\!v_2^4
+{\lambda_{345}\over4}~\!v^2_1v^2_2~\!.\label{eqn:Vtree}
\end{eqnarray}
In this letter
we are interested in the IDM phase in which the electroweak symmetry is
broken by the (real) nonzero vacuum expectation value (VEV) $v_1=v$
while the second doublet $\Phi_2$ does not develop any VEV ($v_2=0$).
This requires $\lambda_1>0$ and $m^2_{11}<0$; the role of the Higgs 
particle is then played by $h^0$.
This vacuum preserves  the $\mathbb{Z}_2$ symmetry and the lightest of
the spinless particles $H^0$, $A^0$ and $H^\pm$, being odd under unbroken
$\mathbb{Z}_2$, is stable. If electrically neutral, it can
therefore constitute the Dark Matter. In the IDM vacuum the (tree-level)
masses of the physical spinless particles which we choose as the
independent variables are given by
\begin{eqnarray}
&&M^2_{h^0}=m^2_{11}+{3\over2}~\!\lambda_1~\!v^2_{\rm tree}=-2m^2_{11}
=\lambda_1v^2_{\rm tree},\phantom{aaaaaaa}\nonumber\\
&&M^2_{H^0}=m^2_{22}+{1\over2}~\!\lambda_{345}~\!v^2_{\rm tree},
\phantom{aaaaaa}\nonumber\\
&&M^2_{H^+}=m^2_{22}+{1\over2}~\!\lambda_3~\!v^2_{\rm tree},
\phantom{aaaaaaa}\label{eqn:masses}\\
&&M^2_{A^0}=m^2_{22}+{1\over2}~\!(\lambda_{345}-2\lambda_5)v^2_{\rm tree}.
\nonumber
\end{eqnarray}
The value of $v_{\rm tree}=246$~GeV is fixed by the Fermi constant $G_F$.
For the remaining six parameters we choose the four physical masses
(\ref{eqn:masses}), and parameters $\lambda_2$
and $\lambda_{345}\equiv\lambda_3+\lambda_3+\lambda_5$. We also take
$\lambda_5$ real and negative\footnote{Any phase factor of $\lambda_5$ can
be removed by a suitable redefinition of the fields. For real $\lambda_5$
the $H^0$ particle corresponds to the real part of the lower component of
$\Phi_2$; the redefinition needed to make $\lambda_5<0$ out of $\lambda_5>0$
amounts to multiplying $\Phi_2$ by $i$, that is, to interchanging $H^0$ and
$A^0$ \cite{GKKS}.}
so that $H^0$ is lighter than $A^0$. With $\lambda_1>0$, $m^2_{11}<0$ and
positive masses squared of all spinless particles the IDM minimum
$v_1^2=v^2_{\rm tree}=-2m^2_{11}/\lambda_1\neq0$, $v_2=0$ is at least a local
minimum\footnote{Owing to the $\mathbb{Z}_2$ symmetry the derivative
with respect to $v_2$ automatically vanishes at this point.}
of the potential (\ref{eqn:Vtree}).  The absolute stability
(boundedness of the potential from below) along the directions preserving
the $U(1)_{\rm EM}$ symmetry requires in addition $\lambda_2>0$ and
$|\lambda_{345}|<\sqrt{\lambda_1\lambda_2}$ while boundedness of the
potential along the electromagnetic symmetry breaking directions requires
also that $|\lambda_3|<\sqrt{\lambda_1\lambda_2}$. However if the IDM is
treated as an effective theory valid only up to some high cut-off scale
$\Lambda$, boundedness of the potential from below for arbitrarily large
field values is not a physical requirement: it
is sufficient to require that the electroweak symmetry breaking vacuum
be the deepest minimum\footnote{Barring the possible metastability of
the present phase of the Universe \cite{stru}.}
in the domain in which $|v_{1,2}|\simlt\Lambda$.

In the following we fix the SM-like Higgs boson $h^0$ mass $M_{h^0}$ to
125 GeV, consistently with the recent indications from the LHC \cite{LHC}.
There are then three possible ranges of the $H^0$ mass ($M_{H^0}$) for which
the right relic density of these particles can be generated during
the evolution of the Universe \cite{laura,su,dsok}: i) $M_{H^0}\simgt$TeV, ii)
$M_{H^0}\sim40\div80$ GeV, and iii) $M_{H^0}\simlt8$ GeV. We will not
consider the first possibility: with very heavy $A^0$ and $H^\pm$
($M_A\sim M_{H^\pm}\simgt M_{H^0} \simgt$ 1 TeV) the temperature
properties of the potential should be identical to the ones of the Standard
Model potential (decoupling!), that is the electroweak phase transition would
be of second order. With the DM particle mass $M_{H^0}$ smaller than $M_W$
(scenarios ii) and iii)),
too light $A^0$ and $H^\pm$ would not produce in the high temperature
expansion (\ref{eqn:HighTexp}) a cubic term of a magnitude necessary for
a sufficiently strong electroweak phase transition. As in \cite{CHNESEZH}
we concentrate therefore on heavy $A^0$ and $H^\pm$. The electroweak
precision data require then the masses of these particles to be 
degenerate - otherwise the contribution of the extended scalar
sector to the Peskin-Takeuchi $T$ and $S$ parameters would be too large.
We thus take $M_{H^0}\ll v_{\rm tree}$ and $M_{A^0}\approx M_{H^\pm}>M_{h^0}$
\cite{su,dsok}.
This mass configuration allow also to satisfy the existing collider limits
\cite{LUGUED,ABDELPHI}.

Because
\begin{eqnarray}
\lambda_5=(M^2_{H^0}-M^2_{A^0})/v^2_{\rm tree},\phantom{aaaa}
\lambda_3=2(M^2_{H^\pm}-M^2_{H^0})/v^2_{\rm tree}+\lambda_{345},
\end{eqnarray}
$A^0$ and $H^\pm$ significantly heavier than $H^0$ imply large values of the
couplings
$\lambda_3$ and $\lambda_5$ so that unitarity of the tree-level scattering
amplitudes in the scalar sector imposes an upper bound of $\sim700$~GeV on the
$A^0$ and $H^\pm$  masses \cite{GOKR}. Alternatively, as done in
\cite{CHNESEZH}, the coupling $\lambda_3$ (and consequently also $\lambda_5$
if $A^0$ and $H^\pm$ are to be degenerate) can be constrained by imposing a
(to a large extent arbitrary) bound on the growth of $\lambda_1$
with the renormalization scale. However, as we show in the next
section, the requirement that the minimum $v_1=v_{\rm tree}$, $v_2=0$ be the
absolute minimum (in the domain $|v_{1,2}|\simlt\Lambda$)  leads to the
bound $M_{A^0}, M_{H^\pm}\simlt440$ GeV. This result, obtained  after
taking into account quantum corrections
to the zero-temperature effective potential, weakly depend on the
values of $\lambda_2$ and $\lambda_{345}$.

\section{One-loop effective potential at T=0}

The one-loop effective potential $V_{\rm eff}(v_1,v_2)$
 is given in the Landau gauge by the standard formula
\begin{eqnarray}
V^{(1L)}_{\rm eff}=V_{\rm tree}+{1\over64\pi^2}\sum_{\rm fields}C_s\left\{
{\cal M}^4_s\left(\ln{{\cal M}^2_s\over4\pi\mu^2}-{3\over2}
+{2\over d-2}-\gamma_{\rm E}\right)\right\}+{\rm CT},\label{eqn:VCW}
\end{eqnarray}
where ${\cal M}^2_s(v_1,v_2)$ are field dependent masses squared
(eigenvalues of the appropriate matrices),
$C_s=(-1)^{2s}(2s+1)g_s$ accounts for the number of states
and $V_{\rm tree}$ is given by (\ref{eqn:Vtree}).
The sum over fields does not include ghost but does include the would-be
Goldstone bosons. We specify the counterterms CT by imposing the following
renormalization conditions.

Firstly we require that the first derivative of $V_{\rm eff}$ with respect
to $v_1$ vanishes at $v_1=v_1^{\rm tree}=-2m^2_{11}/\lambda_1$. This is
equivalent to fixing the
Lagrangian counterterm linear in the $h^0$ field
\begin{eqnarray}
\delta{\cal L}_{\rm lin}=-[\delta m^2_{11}+m^2_{11}\delta Z_1
+{1\over2}(\delta\lambda_1+2\delta Z_1)(v_1^{\rm tree})^2]~\!v_1^{\rm tree} h^0
\nonumber\\
=-[\delta m^2_{11}-m^2_{11}\delta Z_1-{m^2_{11}\over\lambda_1}
\delta\lambda_1]~\!v_1^{\rm tree} h^0,\phantom{aaaaaaaa}
\end{eqnarray}
($\delta Z_i$, $i=1,2$ are the renormalization constants of the two doublets
$\Phi_i$) so that it cancels the one-loop 1PI
tadpole\footnote{Since the formula
(\ref{eqn:VCW}) is obtained in the dimensional reduction rather than
in the dimension regularization, the tadpole $-i{\cal T}_h$ and the
self-energies $\Sigma_h$ and $\Sigma_H$ must also be computed using
the dimensional reduction.}
$-i{\cal T}_h$ of the field $h^0$.
This automatically ensures that the would-be Goldstone boson propagators
have, in the Landau gauge, poles at $p^2=0$. Next we require that the $h^0$
field propagator has the pole for the tree-level mass-squared with the
residue equal $i$. Together these conditions determine the combinations
($\Sigma_h(p^2)$ is the $h^0$ field self-energy):
\begin{eqnarray}
\delta m^2_{11}+m^2_{11}\delta Z_1=-{1\over2}\left(
3{{\cal T}_h\over v_1^{\rm tree}}-\Sigma_h(M^2_h)+M^2_h\Sigma^\prime_h(M^2_h)
\right),\nonumber\\
\delta\lambda_1+2\lambda_1\delta Z_1=-{\lambda_1\over2 m^2_{11}}
\left({{\cal T}_h\over v_1^{\rm tree}}-\Sigma_h(M^2_h)
+M^2_h\Sigma^\prime_h(M^2_h)\right),
\end{eqnarray}
which renormalize the divergent parts of $V^{(1L)}_{\rm eff}$ proportional
respectively to $v_1^2$ and $v_1^4$. As long as the effective potential
is probed only along the $v_1$ direction these two combinations are all what
is needed; in particular, switching off the $\mathbb{Z}_2$-odd fields
one recovers the effective potential of the SM with all its well known
features (including the Linde-Weinberg \cite{LINWEI} lower bound on
the $h^0$ Higgs boson mass).

\begin{figure}[]
\begin{center}
\centerline{\includegraphics[width=0.5\textwidth]{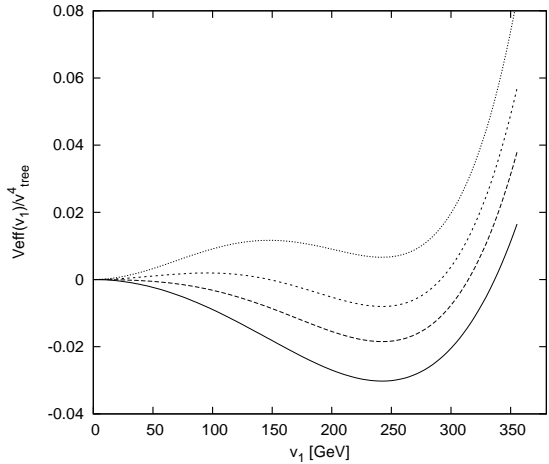}}
\caption{Behaviour of the zero temperature effective potential for
$\lambda_{345}=0.2$, $\lambda_2=0.2$, $M_{h^0}=125$, $M_{H^0}=65$ and
$M_{H^\pm}=M_{A^0}=300$ GeV (solid line), 400 GeV (long-dashed) 450 GeV (
short-dashed) and 500 GeV (dotted).}
\label{fig:pot}
\end{center}
\end{figure}

For the other counterterms needed to renormalize the effective potential there
does not seem to exist equally obvious physical conditions. Any choice of these
counterterms corresponds to some particular definition of the renormalized
couplings $\lambda_2$ and $\lambda_{345}$. The parameter $\lambda_{345}$
determines
the coupling of the SM-like Higgs particle to the DM $H^0$ particles
and the remaining counterterms could in principle
be chosen so that $\lambda_{345}$ is directly related to the physical
$h^0\rightarrow H^0H^0$  decay amplitude. On the other hand
$\lambda_2$
cannot be directly measured in the foreseeable future\footnote{Some
constraints on $\lambda_2$ follow from the DM relic density \cite{dsok}.}
so its precise definition at the loop-level is not important. Here for
simplicity we choose to subtract the divergences of
$V^{(1L)}_{\rm eff}$ proportional to $v_2^4$ and $v_1^2v_2^2$ using the
$\overline{\rm MS}$ scheme. This fixes the combinations
$\delta\lambda_2+2\lambda_2\delta Z_2$ and
$\delta\lambda_{345}+\lambda_{345}(\delta Z_1+\delta Z_2)$. Once the latter
counterterm is fixed the last necessary combination
$\delta m^2_{22}+m^2_{22}\delta Z_2$  is determined by renormalizing the
$H^0$ propagator on-shell. The counterterms $\delta\lambda_3$ and
$\delta\lambda_5$ can be then used to enforce that the tree-level
masses $M_{A^0}$ and $M_{H^\pm}$ remain unchanged by one-loop corrections
(they do not need to be determined explicitly).

\begin{figure}[h]
\centering
\includegraphics[width=10cm]{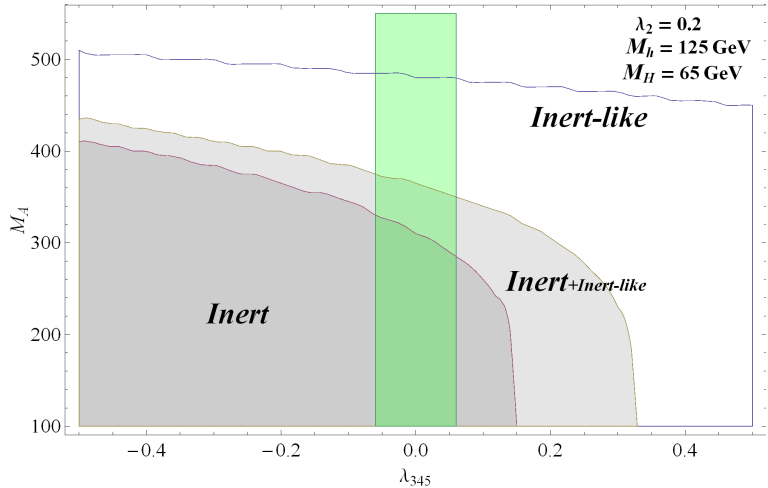}
\caption{Different $T=0$ phases of the model in the ($\lambda_{345},M_{A^0}$)
plane (see the text) for $\lambda_2=0.2$, $M_{h^0=}125$ GeV, $M_{H^0}=65$ GeV
and $M_{H^\pm}=M_{A^0}$. Vertical band corresponds to  the
$\lambda_{345}$-region allowed by the Xenon 100 data. }
\label{fig:wykres1}
\end{figure}

The typical behaviour of the zero-temperature effective potential
$V^{(1L)}_{\rm eff}$
(\ref{eqn:VCW}) along the direction $v_2=0$ for different values of the heavy
$\mathbb{Z}_2$-odd particles masses is shown in fig. \ref{fig:pot}.
It is clear that
for too heavy $H^\pm$ and $A^0$  the electroweak symmetry breaking minimum of
$V^{(1L)}_{\rm eff}$ becomes metastable because while still remaining a local
minimum, it becomes higher than the minimum at $v_1=0$. With two doublets it
is however also possible that the full one-loop $T=0$ potential
develops other deeper minima and this is indeed what happens: as the masses
of the $H^\pm$ and $A^0$ particles grow (remaining almost degenerate) for
fixed value of $\lambda_{345}$ a new minimum of $V^{(1L)}_{\rm eff}$  appears
along the direction $v_2\neq0$, $v_1=0$ and, above some critical value of
$M_{A^0}\approx M_{H^\pm}$, it becomes deeper than the minimum at
$v_1=v_{\rm tree}$, $v_2=0$ whose existence - at least as a local one - is
enforced by our renormalization condition. As illustrated in fig.
\ref{fig:wykres1}, in the plane $(\lambda_{345},~\!M_{A^0})$
one can  distinguish three domains: in
the first one - denoted   ``Inert'' - the minimum
$v_1=v_{\rm tree}$, $v_2=0$ is the global minimum and the only other (local)
minimum can be the one at $v_1=v_2=0$ (cf. the long-dashed line in
fig. \ref{fig:pot}). In the domain
denoted ``Inert$+$inert-like'' the minimum at $v_1=0$, $v_2\neq0$ exists
but the one at  $v_1=v_{\rm tree}$, $v_2=0$ is still the deepest minimum.
Finally, in the domain denoted ``Inert-like'' the minimum at $v_1=0$,
$v_2\neq0$ becomes the deepest one and the inert phase could only exist as
metastable one (and probably would not be reached in the course of the
thermal evolution of the Universe). The upper (almost horizontal) line in
fig.~\ref{fig:wykres1} delimits the region in which the minimum at $v_1=v_2=0$
is deeper than the one at  $v_1=v_{\rm tree}$, $v_2=0$ (i.e. it corresponds to
the metastability of the inert phase illustrated in fig. \ref{fig:pot}).
The central vertical band marked in  fig. \ref{fig:wykres1}
shows the range of the coupling $\lambda_{345}$ still allowed by the
negative results of the XENON 100 experiment \cite{XENON}. (One
should however remember that the coupling $\lambda_{345}$ defined in our
renormalization scheme can differ by
from the effective $h^0H^0H^0$ coupling tested in this experiment.)

Figure  \ref{fig:wykres1} shows that imposing the condition that the
Inert phase of the model be an absolutely stable after including the
one-loop quantum corrections constrains the (degenerate) masses of
$A^0$ and $H^\pm$ to be smaller than $\sim440$~GeV (smaller than
$\sim380$ ~GeV if the XENON 100 limit on $\lambda_{345}$
can be trusted).
Thus, the upper limit imposed on these masses  in \cite{CHNESEZH} by appealing
to a rather at hoc criterion can be replaced by a more physical one. Note,
that the regions marked in fig.~\ref{fig:wykres1} do not change considerably
for $M_{h^0}$ in the range $120\div130$ GeV and are very weakly sensitive to
the DM mass $M_{H^0}\simlt80$ GeV and to the value of $\lambda_2$
(between 0 and 1).

\section{Temperature-dependent effective potential}

In this section we investigate the strength of the electroweak phase
transition predicted by the IDM. Our analysis goes beyond that
of \cite{CHNESEZH}
which was limited by the use of the high temperature expansion, not well
justified for the realistic values of the particle masses.
Concentrating only on the relevant variables allows us to display
the most characteristic details of the electroweak phase transition
in the scenario whose more broad aspects were analyzed in \cite{BOCL}.

The one-loop temperature dependent effective potential is given by \cite{DOJA}
\begin{eqnarray}
V^{(1L)}_T(v_1,v_2)=V^{(1L)}_{\rm eff}(v_1,v_2)+\Delta^{(1L)} V_{T\neq0}(v_1,v_2).
\end{eqnarray}
$V^{(1L)}_{\rm eff}$ has been specified in the preceding section (eq.
\ref{eqn:VCW}) and
\begin{eqnarray}
\Delta^{(1L)} V_{T\neq0}={T^4\over2\pi^2}\sum_{\rm fields}C_s\!
\int_0^\infty\!dx~\!x^2\ln\left[1-(-1)^{2s}\exp\left(-\sqrt{x^2+{\cal M}_s^2/T^2}
\right)\right].\label{eqn:Vtemp}
\end{eqnarray}

For $T^2\gg{\cal M}^2_s$ the contribution of ${\cal M}^2_s$ to
(\ref{eqn:Vtemp}) can be expanded:
\begin{eqnarray}
\left(\Delta^{(1L)} V_{T\neq0}\right)_B=|C_s|\left\{
-{\pi^2\over90}T^4+{1\over24}T^2{\cal M}^2_s-{T\over12\pi}|{\cal M}^3_s|
-{{\cal M}^4_s\over64\pi^2}\!\left(\ln{{\cal M}^2_s\over T^2}-C_B\right)\right\}
\nonumber\\
\left(\Delta^{(1L)} V_{T\neq0}\right)_F=|C_s|\left\{
-{7\pi^2\over720}T^4+{1\over48}T^2{\cal M}^2_s
+{{\cal M}^4_s\over64\pi^2}\!\left(\ln{{\cal M}^2_s\over T^2}-C_F\right)\right\}
\phantom{aaaaaaaaa}\label{eqn:HighTexp}
\end{eqnarray}
($C_B=5.40762$, $C_F=2.63503$). In the opposite limit $T^2\ll{\cal M}^2_s$
one has
\begin{eqnarray}
\left(\Delta^{(1L)} V_{T\neq0}\right)_s
=-|C_s|~\!T^4\left(
{|{\cal M}_s|\over2\pi T}\right)^{3/2}\!\!\left(1+{15\over8}{T\over|{\cal M}_s|}
+\dots\right)\exp\left(-{|{\cal M}_s|\over T}\right),\phantom{a}
\end{eqnarray}
for bosons and fermions alike. In our numerical investigations we perform
the resummation of the higher-order daisy diagrams. For the contribution of
the scalar sector to the temperature-dependent effective potential
this is achieved by interpreting the $T^2$ dependent terms in the expansion
(\ref{eqn:HighTexp}) as the corrections to the Lagrangian mass parameters
$m^2_{ii}$ $i=1,2$: $m_{ii}^2\rightarrow m^2_{ii}(T)\equiv m_{ii}^2+c_iT^2$ where
\begin{eqnarray}
c_1={3\lambda_1+2\lambda_3+\lambda_4\over12}
+{3g^2+g^{\prime2}\over16}+{g_t^2+g_b^2\over4}~,\nonumber\\
c_2={3\lambda_2+2\lambda_3+\lambda_4\over12}
+{3g^2+g^{\prime2}\over16}~,\phantom{aaaaaaaa}
\end{eqnarray}
($g$, $g^\prime$ and $g_t$, $g_b$ are the gauge and Yukawa couplings,
respectively)
and using $m^2_{ii}(T)$ obtained in this way to calculate the field
dependent masses squared ${\cal M}^2_s$ which are reinserted back into the
formula (\ref{eqn:Vtemp}). For the contribution of the gauge boson
sector we follow the prescription of \cite{CARR}. Unlike the authors
of \cite{CLIKAITR}, in the zero-temperature part $V^{(1L)}_{\rm eff}$
of the potential we use the temperature-independent masses squared
${\cal M}_s^2$; inserting  temperature dependent masses into 
$V^{(1L)}_{\rm eff}$ would amount to generating inadmissible UV divergences 
depending on temperature $T$.

\begin{figure}[]
\centering
\includegraphics[width=0.49\textwidth]{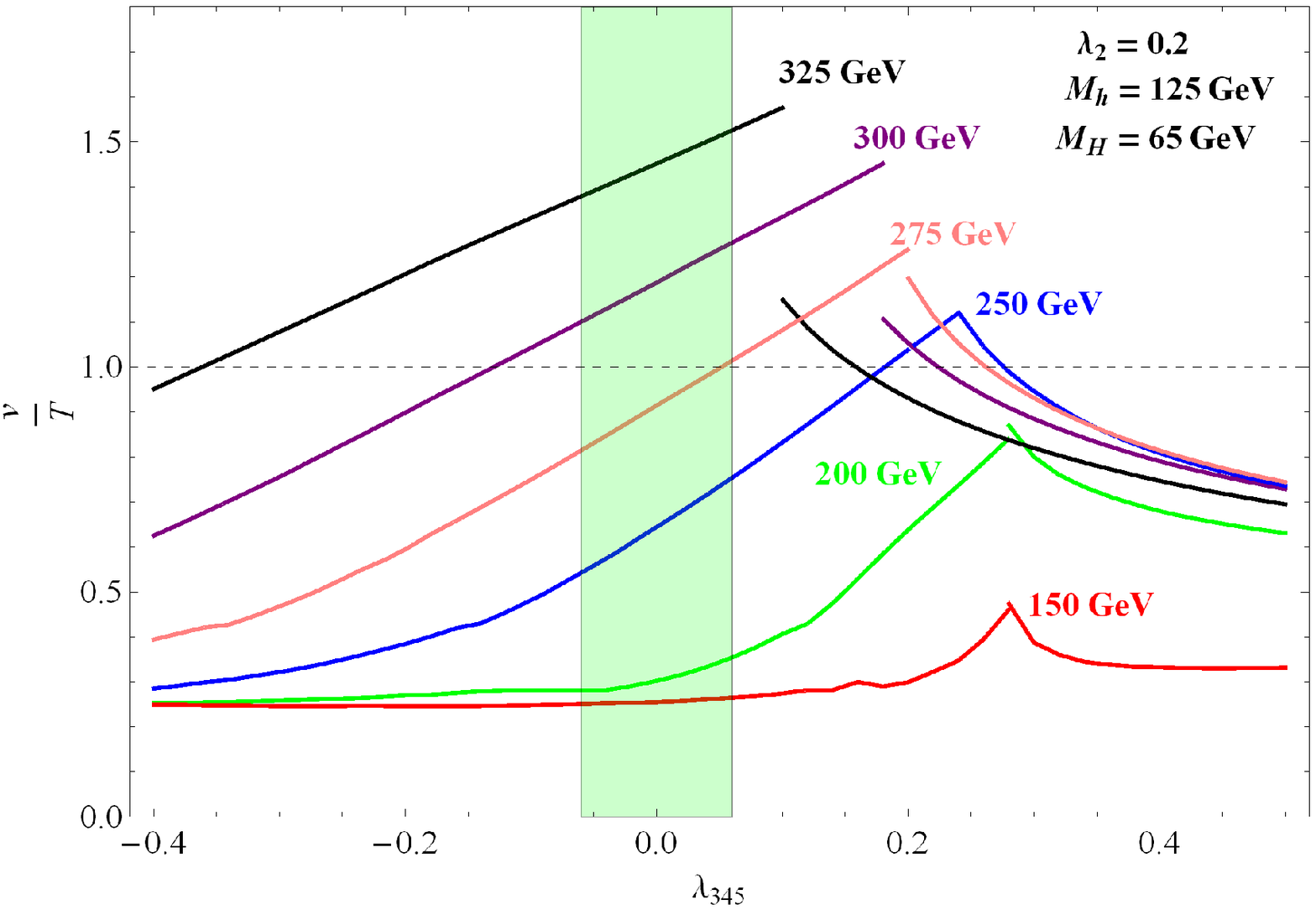}
\includegraphics[width=0.49\textwidth]{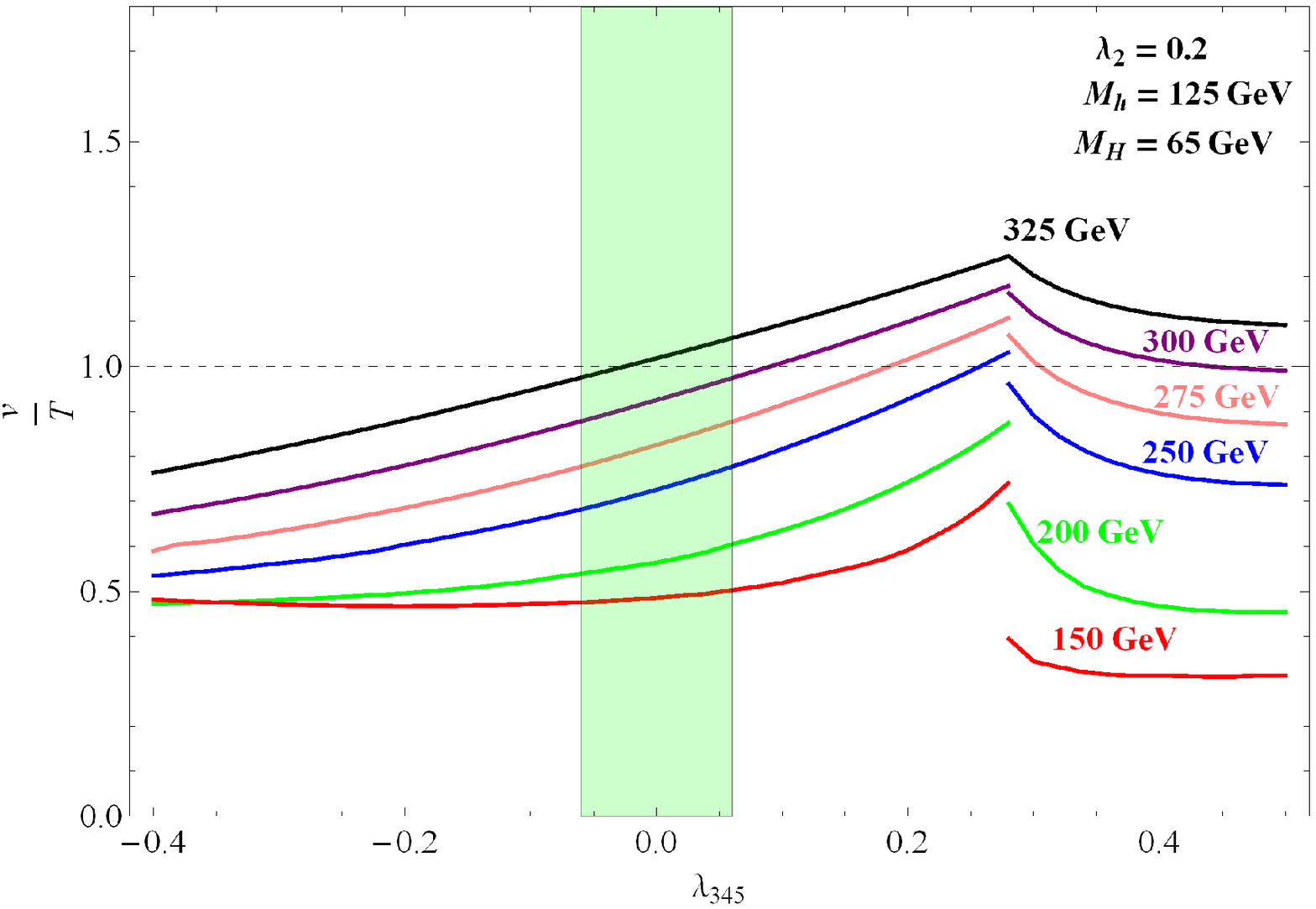}
\caption{The strength of the electroweak phase transition
for $M_{h^0}=125$~GeV, $M_{H^0}=65$~GeV, $\lambda_2=0.2$ as a function
of the coupling $\lambda_{345}$ for different values of $M_{A^0}=M_{H^\pm}$.
Left panel: with  the zero-temperature potential $V_{\rm eff}^{(1L)}$ included.
Right panel: without $V_{\rm eff}^{(1L)}$. The shaded vertical band
corresponds to the region allowed by the Xenon 100 data.}
\label{wykres3}
\end{figure}

We have probed the potential $V^{(1L)}_T(v_1,v_2)$ as a function of
two variables $v_1$ and $v_2$ looking for minima appearing away from
the origin $v_1=v_2=0$ as the temperature $T$ is lowered.
The critical temperature $T_{EW}$ is defined as the one for which the value
of $V^{(1L)}_T$ at the new minimum is equal to its  value for $v_1=v_2=0$.
The measure of the strength of the phase transition is given by the ratio
$v(T_{EW})/T_{EW}$, where $v=\sqrt{v^2_1+v^2_2}$ determines the $W$ boson mass.

\begin{figure}[]
\centering
\includegraphics[width=8.5cm]{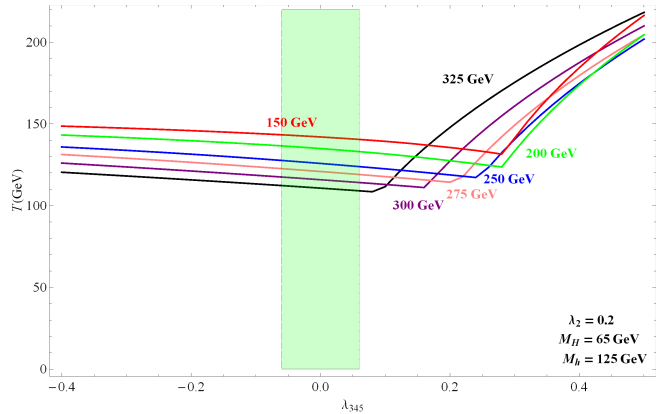}
\caption{$T_{EW}$ as a function of $\lambda_{345}$ for the same parameters
as in fig. \ref{wykres3}.}
\label{wykres4}
\end{figure}

Even if the free parameters of the model correspond to the domain ``Inert''
in fig. \ref{fig:wykres1}, thermal evolution of the system can be quite
complicated \cite{GKKS, dsok}: it can first go to a minimum other than
$v_1\neq0$, $v_2=0$ and jump to it only after further cooling. The
electroweak phase transition consists then of two consecutive transitions
of different strengths. We have found that in the considered scenario this
indeed can happen: for
$\lambda_{345}$ larger than some critical value (which depends on the $A^0$
and $H^\pm$ masses) there appears first the minimum at $v_1=0$,
$v_2\neq0$. This is seen in fig. \ref{wykres3} where we show
$v(T_{EW})/T_{EW}$ as a function of $\lambda_{345}$ for several values of the
$A^0$  and $H^\pm$ masses.  The left panel of this figure shows result
obtained using the temperature-dependent potential with the Coleman-Weinberg
term while the right one - without this term. To the left of the
discontinuities (or cusps) of the curves
the electroweak phase transition occurs in one step:
$(0,0)\rightarrow (v_1,0)$. The discontinuities (cusps) mark the
critical values of $\lambda_{345}$ for which the system goes first to
the minimum $(0,v_2)$. However, for the  allowed by Xenon 100 values of
$\lambda_{345}$ (vertical bands)
the phase transition occurs in one step and  for
$M_{A^0}\approx M_{H^\pm}$ in a rather narrow window between $\sim275$~GeV
and $\sim380$~GeV (the upper limit of the ``Inert'' domain allowed by
the XENON 100 experiment in
fig.~\ref{fig:wykres1}) it is sufficiently strong to allow for electroweak
baryogenesis. The corresponding values of the temperatures $T_{\rm EW}$
are shown in fig.~\ref{wykres4}. The turnover points correspond
to the changes in the character of the symmetry
breaking minimum discussed above. For the considered masses of the scalars
the temperatures of the EW transition lay between 100 and 150 GeV.

Comparison of the left and right panels of fig. \ref{wykres3}  illustrates
the important impact of the zero-temperature potential $V_{\rm eff}^{(1L)}$ 
(\ref{eqn:VCW}) on the strength of the phase transition.\footnote{The
contribution of $V_{\rm eff}^{(1L)}$ was also taken into account in 
\cite{BOCL} but was omited in \cite{CHNESEZH}).}
It is clear that neglecting $V_{\rm eff}^{(1L)}$  underestimates the value
of $v(T_{EW})/T_{EW}$. The second effect of $V_{\rm eff}^{(1L)}$ is the
welcome shift of the maximal value of this ratio to the left, closer to
the band of $\lambda_{345}$ values allowed by the XENON 100 results.

The values of  $v(T_{EW})/T_{EW}$ shown in fig. \ref{wykres3}
depend rather weakly on the coupling $\lambda_2$ and the masses
$M_{h^0}$ and $M_{H^0}$ in the considered ranges. For instance
changing $M_{h^0}$ from 120 GeV to 130 GeV decreases $v(T_{EW})/T_{EW}$
by a factor $\sim0.85$. The ratio $v(T_{EW})/T_{EW}$ also mildly 
increases with decreasing mass of the DM particle.

\section{Conclusions}

We have reconsidered the electroweak phase transition predicted by the
inert doublet model with a stable DM candidate. For the DM particle below 
the electroweak scale the ratio $v(T_{EW})/T_{EW}$ which determines the
strength of the phase transition depends mainly on the coupling
$\lambda_{345}$ and the masses of the additional spinless particles $A^0$
and $H^\pm$. For $v(T_{EW})/T_{EW}\simgt1$ these states must be sufficiently
heavy. They are then forced to be highly degenerate
(to satisfy the constraints from the electroweak precision data)
and, as we have found, their (common) mass is strongly bounded from above
by the requirement that the  inert vacuum is reached at the end of the
thermal evolution (i.e. that this minimum is the deepest one).
If the XENON 100 constraint on the effective $h^0H^0H^0$
coupling of the DM to the SM-like Higgs boson can be applied directly
to $\lambda_{345}$, one concludes that the portion of the
IDM parameter space in which it can predict the right relic
density of DM particles density and a sufficiently strong electroweak
phase transition is rather limited. In particular, the masses of the
$H^\pm$ and $A^0$ states are constrained to a narrow window $275-380$ GeV.
We also stress that it is the zero temperature part of the potential
which allows to reconcile the requirement $v(T_{EW})/T_{EW}\simgt1$
with the constraints following from stability and the XENON 100 results.

Finally, if the necessary additional source of CP violation is due to the
operator (\ref{eqn:CPVOperator}), the mechanism producing the excess of
baryons should be due to the so-called local baryogenesis for
which the generated value $n_B/s$ of the baryon number to entropy ratio
can reliably be estimated only in the quasi-static regime (thick, slowly
moving walls of the bubbles of the new phase) \cite{DIHUSISU,ZHYOU}; in the
opposite regime of fast change the value $n_B/s$ is
rather hard to estimate even if the scale $\Lambda$
in (\ref{eqn:CPVOperator}) is known \cite{DIHUSISU,LURATR,TROD}.
It is also interesting to note that the operator (\ref{eqn:CPVOperator})
would contribute to the nowadays very important decay 
$h^0\rightarrow\gamma\gamma$:
\begin{eqnarray}
\Gamma(h^0\rightarrow\gamma\gamma)=
{\alpha^3_{\rm EM}G_{\rm F}M^3_{h^0}\over128\pi^3}\left\{\left|A_{\rm SM}+
A_{H^+}\right|^2+\left|4c_W~\!{v_{\rm tree}^2\over\Lambda^2}\right|^2\right\},
\end{eqnarray}
where $A_{\rm SM}\approx-6.5$ for $M_{h^0}=125$~GeV, $A_{H^+}$ is the $H^+$
contribution \cite{PO} and where as in \cite{ZHYOU} we have written
the coefficient $c_1$ of the operator (\ref{eqn:CPVOperator}) as
$c_W~\!g^2/8\pi^2$. With $c_W\sim0.1\div1$ \cite{ZHYOU} and $\Lambda$
in the TeV range this contribution is small but potentially distinguishable
(though not at the LHC)
as the two photons originating from the interaction (\ref{eqn:CPVOperator})
are polarized differently compared to the  photons originating from the
ordinary loop-induced coupling.

\vskip0.5cm

{\bf Acknowledgments.} We thank D. Soko\l owska and M. Carrington for
valuable discussions. The work of G.G. and M.K. was partly supported by
the Polish Ministry of Science and Higher Education Grant N202 230337.


\newpage
\begin{thebibliography}{99}

\bibitem{JANS} K. Kajantie {\it et al.} {\sl Phys. Rev. Lett.} {\bf 77}
               (1996) 2887 hep-ph/9605288];
               {\sl Nucl. Phys.}, {\bf B493} (1997) 413 [hep-lat/9612006];
               M. Gurtler, E.-M. Ilgenfritz and A. Schiller, {\bf Phys. Rev.}
               {\bf D56} (1997) 3888
               K. Jansen {\sl Nucl. Phys. Proc. Suppl.}, {\bf 47} (1996) 196.

\bibitem{LEP} LEP Collaborations, LEP Electroweak Working Group, SLD
              Electroweak Group and SLD Heavy Flavor Group,
              Report CERN-EP/2003-091, LEP-EWWG/2003-02, hep-ex/0312023.
\bibitem{LHC} G. Aad {\it et al.} (ATLAS Collaboration), {\sl Phys. Lett.}
                     {\bf B710} (2012) 49 [arXiv:2012.1408 [hep-ex]],\\
              S. Chatrchyan {\it et al.} (CMS Collaboration), 
              {\sl Phys. Lett.} {\bf B710} (2012) 26 
              [arXiv:2012.1488 [hep-ex]].

\bibitem{TROD} M. Trodden, {\sl Rev. Mod. Phys.} {\bf 71} (1999) 1463 
               [hep-ph/9803479].

\bibitem{WAGNERY} M.~S.~Carena, M.~Quiros and C.~E.~M.~Wagner,
                  {\sl Phys. Lett.} {\bf B380} (1996) 81 [hep-ph/9603420],
                  M.~Carena, G.~Nardini, M.~Quiros and C.~E.~M.~Wagner,
                  {\sl Nucl. Phys.} {\bf B812} (2009) 243
                  [arXiv:0809.3760 [hep-ph]].

\bibitem{BOKUSH} A.I. Bochkarev, S.V. Kuzmin and M.E. Shaposhnikov,
                {\sl Phys. Lett.} {\bf B244} (1990) 275.

\bibitem{TURZAD} N.~Turok and J.~Zadrozny,  {\sl Nucl. Phys.} {\bf B358}
                 (1991) 471,
                {\sl Nucl. Phys.} {\bf B369} (1992) 729.

\bibitem{SCHAP} M.E.~Shaposhnikov, {\sl JETP Lett.}  {\bf 44} (1986) 465
                [{\sl Pisma Zh. Eksp. Teor. Fiz.}  {\bf 44} (1986) 364];
                {\sl Nucl. Phys.}  {\bf B287}, 757 (1987).

\bibitem{LACAR} D.~Land and E.D.~Carlson,  {\sl Phys. Lett.} {\bf B292}
               (1992) 107 [arXiv:hep-ph/9208227].
\bibitem{LASPZA} A.B. Lahanas, V.C. Spanos and V. Zarikas, {\sl Phys. Lett.}
                 {\bf B472} (2000) 119 [hep-ph/9812535].

\bibitem{CLILEM} J.M.~Cline and P.A.~Lemieux, {\sl Phys. Rev.} {\bf D55}
                 (1997), 3873  [arXiv:hep-ph/9609240].
\bibitem{KAYASE} S.~Kanemura, Y.~Okada and E.~Senaha, {\sl Phys. Lett.}
                 {\bf B606} (2005), 361 [arXiv:hep-ph/0411354].

\bibitem{FROHUSE} L.~Fromme, S.J.~Huber and M.~Seniuch, {\sl JHEP}
                  {\bf 0611} (2006), 038 [arXiv:hep-ph/0605242].

\bibitem{MA} E.~Ma, {\sl Phys. Rev.} {\bf D73} 077301 (2006) [hep-ph/0601225].

\bibitem{BAHARY} R. Barbieri, L.J. Hall and V.S. Rychkov
                 {\sl Phys. Rev.} {\bf D74}:015007 (2006) [hep-ph/0603188].

\bibitem{BFLRSS} G.~C.~Branco, P.~M.~Ferreira, L.~Lavoura, M.~N.~Rebelo, 
                 M.~Sher and J.~P.~Silva, {\sl Phys. Rept.}  {\bf 516}, 
                 1 (2012) [arXiv:1106.0034 [hep-ph]].

\bibitem{GIIVKA} I.F.~Ginzburg, I.P.~Ivanov, K.A.~Kanishev, {\sl Phys. Rev.}
                 {\bf D81}:085031 (2010) [arXiv:0911.2383 [hep-ph]].

\bibitem{GKKS} I.F. Ginzburg, K.A. Kanishev, M. Krawczyk, D. Soko\l owska,
                {\sl Phys. Rev.} {\bf D82}, 123533 (2010); PoS
                {\bf QFTHEP2010} (2010) 067.
\bibitem{dsok} D.~Sokolowska, arXiv:1104.3326 [hep-ph],
              {\sl Acta Phys. Polon.}  {\bf B42}, 2237 (2011)
              [arXiv:1112.2953 [hep-ph]].

\bibitem{GIL} G.Gil,  Master Thesis, Faculty of Physics, University of Warsaw, 
              Badanie przejsc fazowych w Inert Doublet Model, August 2011

\bibitem{KOSKA} A.~Kozhushko and V.~Skalozub, {\sl Ukr. J. Phys.}
                {\bf 56} (2011) 431 [arXiv:1106.0790 [hep-ph]].

\bibitem{CLIKAITR} J.M.~Cline, K.~Kainulainen and M.~Trott, {\sl JHEP}
                {\bf 1111} (2011) 089 [arXiv:1107.3559 [hep-ph]].

\bibitem{laura} L.~Lopez Honorez, E.~Nezri, J.~F.~Oliver and M.~H.~G.~Tytgat,
                {\sl JCAP} {\bf 0702}, 028 (2007) [hep-ph/0612275];
                L.~Lopez Honorez and C.~E.~Yaguna, {\sl JHEP} {\bf 1009}, 
                046 (2010) [arXiv:1003.3125 [hep-ph]].
                L.~Lopez Honorez and C.~E.~Yaguna, {\sl JCAP} {\bf 1101}, 
                002 (2011) [arXiv:1011.1411 [hep-ph]].

\bibitem{su} E.~M.~Dolle and S.~Su, {\sl Phys. Rev.}  {\bf D80}, 055012 
             (2009) [arXiv:0906.1609 [hep-ph]].

\bibitem{CHNESEZH} T.A.~Chowdhury, M.~Nemevsek, G.~Senjanovic and Y.~Zhang,
                   {\sl JCAP} {\bf 1202} (2012) 029 [arXiv:1110.5334 
                   [hep-ph]].

\bibitem{BOCL} D.~Borah and J.M.~Cline, arXiv:1204.4722 [hep-ph].

\bibitem{DIHUSISU} M.~Dine, P.~Huet, R.L.~Singleton and L.~Susskind,
                   {\sl Phys. Lett.}   {\bf B257} (1991) 351,
                   {\sl Nucl. Phys.}  {\bf B375} (1992) 625.

\bibitem{XENON} E.~Aprile {\it et al.}  (XENON100 Collaboration),
                          {\sl Phys. Rev. Lett.}  {\bf 105} (2010) 131302
                          [arXiv:1005.0380 [astro-ph.CO]].

\bibitem{BAFESA} A.~Barroso, P.~M.~Ferreira and R.~Santos, {\sl Phys. Lett.}
                 {\bf B632}, 684 (2006) [hep-ph/0507224].

\bibitem{stru} J.~Elias-Miro, J.~R.~Espinosa, G.~F.~Giudice, G.~Isidori, 
               A.~Riotto and A.~Strumia,  {\sl Phys. Lett.}  {\bf B709}, 
               222 (2012) [arXiv:1112.3022 [hep-ph]];
               G.~Degrassi, S.~Di Vita, J.~Elias-Miro, J.~R.~Espinosa, 
               G.~F.~Giudice, G.~Isidori and A.~Strumia, arXiv:1205.6497 
               [hep-ph].

\bibitem{LUGUED} E.~Lundstrom, M.~Gustafsson and J.~Edsjo, {\sl Phys. Rev.} 
                 {\bf D79}, 035013 (2009) [arXiv:0810.3924 [hep-ph]];
                 M.~Gustafsson,  PoS CHARGED {\bf 2010}, 030 (2010)
                 [arXiv:1106.1719 [hep-ph]].

\bibitem{ABDELPHI} J.~Abdallah {\it et al.} [DELPHI Collaboration],
                   {\sl Eur. Phys. J.} {\bf C34}, 399 (2004) [hep-ex/0404012].

\bibitem{GOKR} B.~Gorczyca and M.~Krawczyk, {\sl Acta Phys. Polon.}
               {\bf B42} (2011) 2229 [arXiv:1112.4356 [hep-ph]];
               arXiv:1112.5086 [hep-ph].

\bibitem{DOJA} L.~Dolan i R.~Jackiw, {\sl Phys. Rev.} {\bf D9} (1974), 3320;
               S.~Weinberg, {\sl Phys. Rev.} {\bf D9} (1974), 3357.

\bibitem{LINWEI} A.D.~Linde, {\sl Pis'ma w Zh. Eksp. Teor. Fiz.} {\bf 23}
                 (1976) 73;
                 S.~Weinberg, {\sl Phys. Rev. Lett.} {\bf 36} (1976) 294.

\bibitem{CARR} M.E. Carrington, {\sl Phys. Rev.} {\bf D45} (1992), 2933.

\bibitem{LURATR} A.~Lue, K.~Rajagopal and M.~Trodden, {\sl Phys. Rev.}
                 {\bf D56} (1997) 1250 [hep-ph/9612282].

\bibitem{ZHYOU} X. Zhang and B.L. Young, {\sl Phys. Rev.} {\bf D49} 
                (1994) 563 [hep-ph/9309269].

\bibitem{PO} P. Posch, {\sl Phys. Lett.} {\bf B696} (2011) 447
             [arXiv:1001.1759 [hep-ph]];
             A.~Arhrib, R.~Benbrik and N.~Gaur,{\sl Phys. Rev.} {\bf D85} 
             (2012) 095021 [arXiv:1201.2644 [hep-ph]].

\end{thebibliography}
\end{document}